# Boosting the SiN nonlinear photonic platform with transition metal dichalcogenide monolayers

Vincent Pelgrin[1,2*], Yuchen Wang[2], Jonathan Peltier[1], Carlos Alonso-Ramos[1], Laurent Vivien[1], Zhipei Sun[2,3], and Eric Cassan[1*]

[1] *Université Paris-Saclay, CNRS, Centre de Nanosciences et de Nanotechnologies, 91120, Palaiseau, France.*

[2] *Department of Electronics and Nanoengineering, Aalto University, P.O. Box 13500, FI-00076 Aalto, Finland*

[3] *QTF Centre of Excellence, Department of Applied Physics, Aalto University, FI-00076 Aalto, Finland\**

*Corresponding author: vincent.pelgrin@universite-paris-saclay.fr, eric.cassan@universite-paris-saclay.fr*

**Abstract**

In the past few years, we have witnessed an increased interest in the use of 2D materials for the realization of hybrid photonic nonlinear waveguides. Although graphene has attracted most of the attention, other families of 2D materials such as transition metal dichalcogenides have also shown promising nonlinear performances. In this work, we propose a strategy for designing silicon nitride waveguide structures embedded with $MoS_2$ for nonlinear applications. The transverse geometry of the hybrid waveguide's structure is optimized for high third-order nonlinear effects using optogeometrical engineering and multiple layers of $MoS_2$. Stacking multiple monolayers, results in an improvement of 2 orders of magnitude in comparison with standard silicon nitride waveguides. The performance of the hybrid waveguides is then investigated in terms of four-wave mixing enhancement in micro-ring resonator configurations. A -6.3 dB signal/idler conversion efficiency is reached around 1.55µm wavelength for a 5 mW pumping level.

Silicon photonics is a very promising platform in many aspects, and especially due to its possible CMOS compatibility. However, integrated silicon nonlinear optical applications at telecom wavelengths e.g 1530 nm - 1560 nm (C-band), are met with a few difficulties. Although silicon is itself a very non-linear material (Kerr index $n_2 \simeq 5.10^{-18} m^2 W^{-1}$[1]), its band structure induces two photon absorption (TPA), thus strongly counterbalancing the optical Kerr nonlinearity of the material. Moreover, the choice of directly compatible materials for photonics is limited. A classical alternative to silicon is silicon nitride (SiN). The absence of TPA in the near IR and the low propagation loss of SiN waveguides (below 1 dB/cm in the near infra-red) make them interesting. Impressive demonstrations of supercontinuum generation [2-4] or frequency comb generation [5-8] have been reported on the SiN/SiO$_2$ platform. Yet, due to its relatively weak Kerr coefficient ($n_2 \simeq 2.4 \times 10^{-19} m^2 W^{-1}$, effective nonlinear coefficient $\gamma \approx 1$ W$^{-1}$m$^{-1}$ [11] against a few hundred W$^{-1}$m$^{-1}$ for Si/SiO$_2$ waveguides [12]), SiN prompts the use of high power pumps in order to meet the high power thresholds required to induce appreciable nonlinear effects (from several hundreds of mW up to a few Watts in continuous-wave power for the generation of broadband frequency combs [5,9,10]). One possible key to unlock possibilities is to turn to the emerging field of two-dimensional (2D) layered materials (e.g. graphene, black phosphorous, transition metal dichalcogenides (TMD)). Recently, promising results of nonlinear frequency conversion using 2D materials have been demonstrated both in free space for second harmonic generation (SHG) and third harmonic generation (THG) [13,30] and in integrated photonics with self-phase modulation (SPM) and four wave mixing (FWM) [14-16]. The idea is thus to exploit this material family in order to solve the limitations faced in standard SiN waveguides by relying on recent breakthroughs made in the improvement of 2D material growth and transfer methods [17-19]

In this context, this paper aims at optimizing the integration strategies of 2D materials for boosting the nonlinear properties of SiN waveguides at telecom wavelengths near 1.5µm, while maintaining TPA-free. The emphasis is first put on the waveguide mode group velocity dispersion profiles (critical for phase matching conditions) and transverse field distributions. In a second step, we report on the optimization of the nonlinear properties of the hybrid waveguides, as well as on the calculation of the overlap between the transverse mode and the 2D material layers. Once nonlinear waveguides are properly designed, the performances of a frequency conversion scheme by considering a ring resonator configuration is chosen as a reference configuration to estimate the nonlinear photonic platform performances [20] The properties of SiN waveguides are well known, both in terms of linear (e.g. dispersion: effective index, group index) and nonlinear properties (nonlinear coefficient $\gamma$). Introducing new materials, even ultra-thin 2D-materials, may affect the linear dispersion properties of the waveguides. It is important to know if this is the case and to be able to quantify it. Moreover, a detailed study of the waveguide dispersion is critical for the design of the optimized structures because of the phase matching requirements in nonlinear optical processes like FWM [21]. A commercial finite difference eigenmodes (FDE) mode solver was used and applied to a TMD-SiN hybrid waveguide in various configurations (see Fig. 1) a)). The effect of one or several monolayers of MoS$_2$ was taken into account by considering the monolayer crystal as a true 2D material modelled thanks to its surface conductivity $\sigma_s$ for the calculation of waveguide modes [22]. The relative permittivity $\epsilon_r$ of the 2D materials was expressed with a sum of N Lorentzian functions. The coefficients $f_k$, $\omega_k$ and $\gamma_k$ are specific to the material and were extracted by fitting experimental data for the most common TMD (MoS$_2$, WS$_2$, MoSe$_2$, WSe$_2$) [23]. $\gamma_k$ stands for the oscillator strength, $\omega_k$ is the resonance frequency and $\gamma_k$ the spectral width of the k$^{th}$ oscillator.

$$\epsilon_r = \frac{\epsilon(\omega)}{\epsilon_0} = 1 + \sum_{k=1}^{N} \frac{f_k}{\omega_k^2 - \omega^2 - i\omega\gamma_k} \tag{1}$$

The complex surface conductivity was then extracted from the dielectric function using the following expression:

$$\epsilon(\omega) = \epsilon_0 \left(1 + \frac{i\sigma_b}{\epsilon_0 \omega}\right) = \epsilon_0 \left(1 + \frac{i\sigma_s}{\epsilon_0 \omega h_{eff}}\right) \tag{2}$$

Related results are shown in Fig. 1) b) [22]. We based the study on SiN /SiO$_2$ waveguides operating at a wavelength range of ~1,0 and 2,0 µm while taking all materials' dispersion into consideration with the Sellmeier equation [24,25]. Transverse electric (TE) like polarization was considered for maximum interaction between the 2D material and the mode [26]. The calculation was thus performed for various waveguide cross-sections. The reference structure (without MoS$_2$) was given for comparison with the hybrid one (with MoS$_2$) differing by the introduction of one or several MoS$_2$ monolayers (and later by an additional oxide cladding).

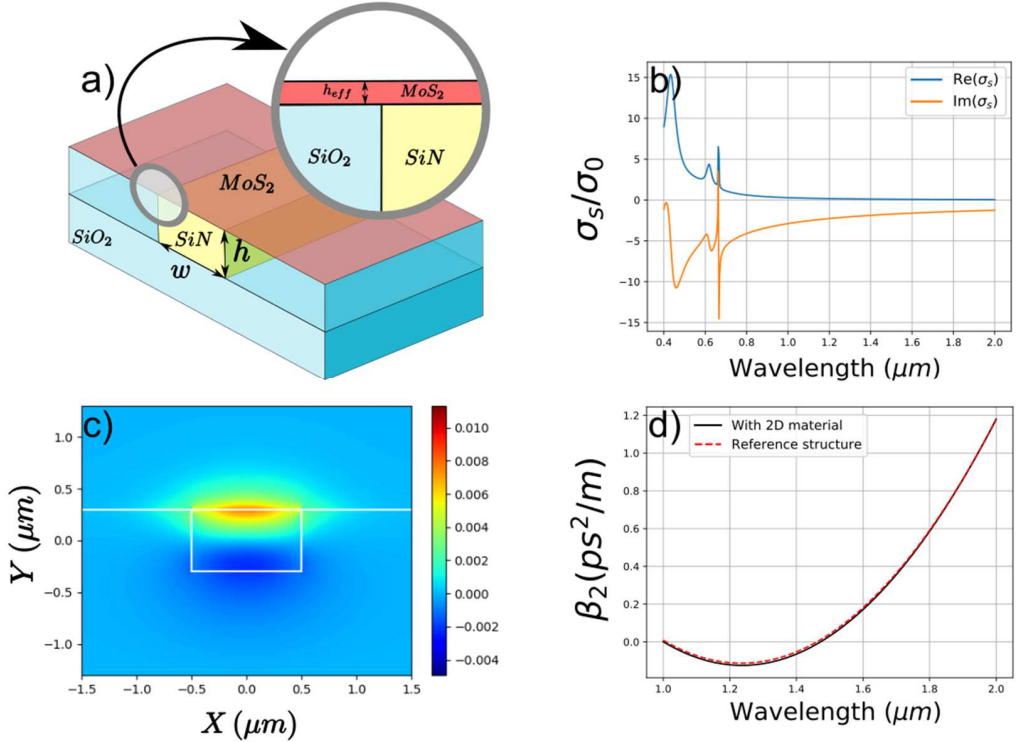

**Fig. 1. a) Studied structure with a 1000 nm wide and 600 nm high waveguide. b) Calculated MoS$_2$ surface conductivity. c) Comparison of the TE mode distribution between a reference structure (without MoS$_2$) and**

a structure covered by monolayer $MoS_2$ at $\lambda$=1,5 µm. The represented field distribution is $E_{2D}$ $E_{ref}$(difference between the two structures' waveguide fields) d) Comparison of the related dispersion profiles.

Starting from a planarized SiN/SiO$_2$ buried strip with a height of $h$ = 600 nm and a width of $w$ = 1000 nm, results are displayed in Fig.1. As visible, the introduction of the 2D material layer has little effect on both the transverse mode distribution (TE single mode) and the waveguide dispersion properties (two orders of magnitude lower difference between the two structure's mode profiles (Fig. 1)c)) with respect to the reference structure field profile: ). Therefore, no extra specific scheme is thus needed for light coupling between the two waveguide families (because modes are matching quite well), thereby simplifying design steps. It also makes the comparison between the reference uncladded SiN waveguides and the hybrid TMD-SiN ones more relevant, as both kinds of waveguides mainly differ only in terms of propagation losses and nonlinearity strength. The literature reports several approaches for the integration of 2D materials [14,26,27]. 2D materials are usually transferred on the chip using wet transfer methods [26,28]. However, once this has been accomplished, other operations can be undertaken. The deposition of an oxide layer such as $Al_2O_3$ or $HfO_2$ on top of the 2D material can be beneficial as it can act as a protective layer.

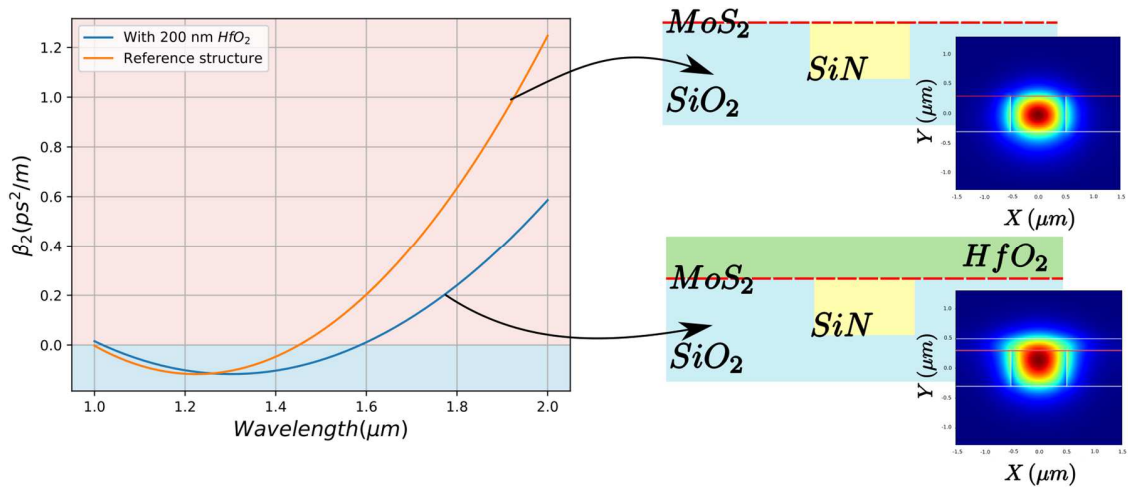

Fig. 2. Comparison between a $HfO_2$ covered structure and a reference structure. The base waveguide is the same as the one displayed in Figure 1.

In addition to this protective action, we propose here to exploit this top cladding layer in another purpose. Because oxides like $Al_2O_3$ or $HfO_2$, that can be grown easily using atomic layer deposition (ALD) [29], present linear refractive indices of 1.7 and 2.07, respectively, the added top cladding layer can improve the passive properties of the waveguide in term of dispersion and directly contribute to the increase of the nonlinear overlap the 2D material layer and the transverse mode (see Fig 2 and Equation (3)). Moreover, this approach does not introduce additional losses since those materials are transparent in the wavelength range of interest. Comparative results between hybrid $MoS_2$-SiN waveguides uncladded and cladded with a 200nm thick $HfO_2$ layer are displayed in Fig 2 for the same base waveguide cross section (w=1000nm and h=600nm) as in Fig 1. A shift in the dispersion profile obviously appears, pushing the mode anomalous dispersion widely in the C band. Furthermore, the spatial overlap factor of the guided quasi-TE mode changes and comes from 0.03% in the absence of $HfO_2$ cladding, to 0.10 %, the high index top layer indeed pulling the mode in the vertical direction. To quantify how this affects the nonlinearities, further studies were performed.

Estimation of the third order nonlinear waveguide performances was achieved as follows [1]. With $e(r, \omega)$ being the transverse electric profile and $\overline{e}(r, \omega)$ its complex conjugate, the effective nonlinear susceptibility of the structure is the result of the integral over the different components when taking into account the effective susceptibility tensor $\chi^{(3)}(r)$ of each considered materials. Only SiN and the 2D layers were considered as being nonlinear, the other material (air, $SiO_2$) nonlinearities being neglected due to their relatively small nonlinear responses and the low mode confinement factor in those materials. Discrepancies in $\chi^{(3)}$ of $MoS_2$ being met in the literature due to different fabrication methods and environmental deposition or growth conditions, either by flake exfoliation or CVD (Chemical Vapor Deposition) growth [13,30], a $n_2$ value reported from in-plane experiments at 1.55 µm wavelength for CVD grown 2D materials was considered as a reference ($n_2$=1.1×10$^{-16}$ m$^2$.W$^{-1}$ [14]). The needed in-plane susceptibility tensor component was then accounted for by considering the effective thickness of the mono-layer $h_{eff}$ (as depicted Fig. 1)a)) while treating the 2D material as a thin 3D material and effectively accounting for the cross-section area occupied by $MoS_2$. The waveguide effective nonlinear susceptibility of the hybrid waveguides was thus estimated as [1]:

$$\Gamma = \frac{A_0 \int_{A_{NL}} \overline{e}(r;\omega)\chi^{(3)}(r)e(r;\omega)\overline{e}(r;\omega)e(r;\omega)dA}{\left(\int_{A\infty} n^2(r)|e(r;\omega)|^2 dA\right)^2} \qquad (3)$$

Here, $A_0$ is the waveguide core area, $A_{NL}$ is the cross section made of all the materials participating to the waveguide effective nonlinearity related to the standard $n_2$ and TPA coefficient ($\beta_{TPA}$) through the following relationship:

$$\frac{\omega}{c} n_2 + \frac{i}{2} \beta_{TPA} = \frac{3\omega}{4\epsilon_0 c^2 n^2} \chi^{(3)} \qquad (4)$$

TPA was further ignored since it is absent from $SiO_2$ and SiN. Concerning $MoS_2$ monolayers, the absence of TPA at $\lambda \approx 1.55$ µm was further recently confirmed by theoretical studies [36]. The waveguide effective nonlinear susceptibility was then used to derive the effective nonlinear Kerr γ (γ ~ 1 W⁻¹m⁻¹ for a standard SiN waveguide) coefficient:

$$\gamma = \frac{3\omega\Gamma}{4\epsilon_0 A_0 v_g^2} \tag{5}$$

, with $\epsilon_0$ being the permittivity in vacuum and $v_g$ the group velocity of the propagating mode. This methodology was used to compare different approaches for the realization and optimization of nonlinear waveguides embedding MoS$_2$ layers.

The main line of the waveguide design was to consider the integration of a variable number of MoS$_2$ monolayers. In order to guarantee a realistic technological approach, the same planarized SiN waveguides as in previous studies were considered. To ensure the MoS$_2$ monolayer character, and thus the large bandgap of the material and the absence of

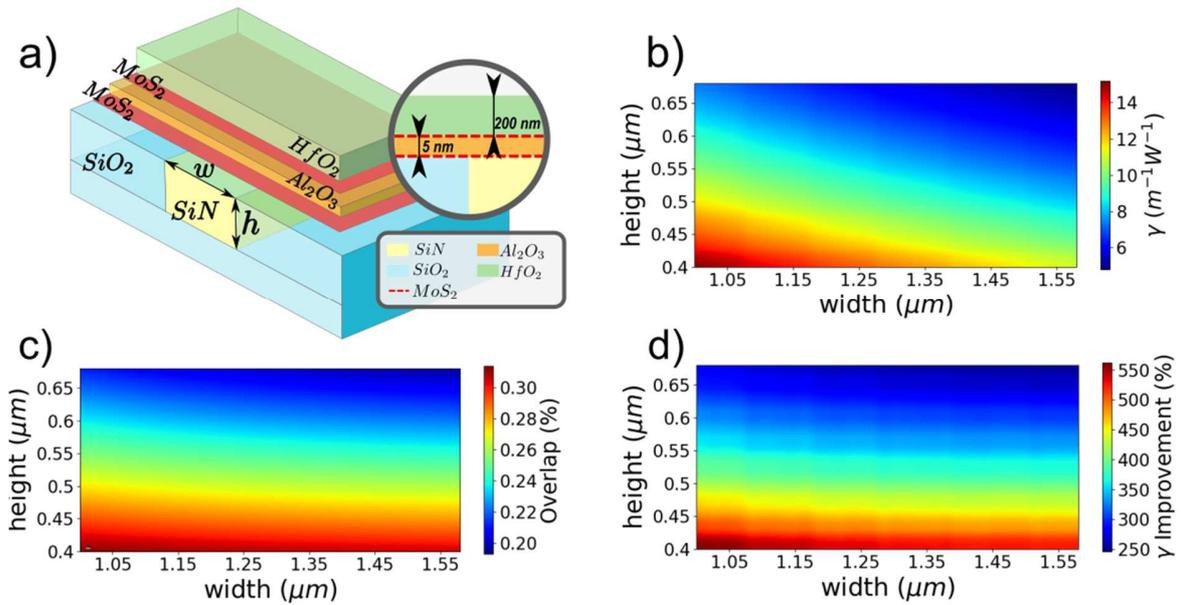

**Fig. 3. a) SiN planarized buried strip covered by two single MoS$_2$ monolayers. b) Nonlinear effective coefficient γ versus the dimensions of the waveguide (height and width). c) Overlap between the 2D materials and the TE mode calculated as the confinement factor of the mode inside the MoS$_2$. d) Improvement factor between the reference structure without anything on top and the structure presented**

TPA [32,36], we introduced an alternation of TMD monolayers separated by Al$_2$O$_3$ spacers of few nm: 5nm was considered as compromise between a safe isolation of single MoS$_2$ mono-sheets and a reasonably thin spacer to minimize the possible thickness of the whole material stack. The proposed approach is compatible with the growth of all materials (TMD, Al$_2$O$_3$ spacers, HfO$_2$) through ALD [29]. Note that the aim of the study is to present a general structure, which later can be improved based on the needs (e.g., Al$_2$O$_3$ and HfO$_2$ can be replaced by other similar insulation materials, e.g., BN). As an intermediate illustrative step, Fig. 3 shows the results obtained for a $MoS_2$ bilayer stack on top of the reference SiN planarized waveguide already considered in Figs. 1 and 2, covered by a 200 nm $HfO_2$ cap to steer the optical mode upward and increase the spatial overlap with the $MoS_2$ 2D mono-layers.

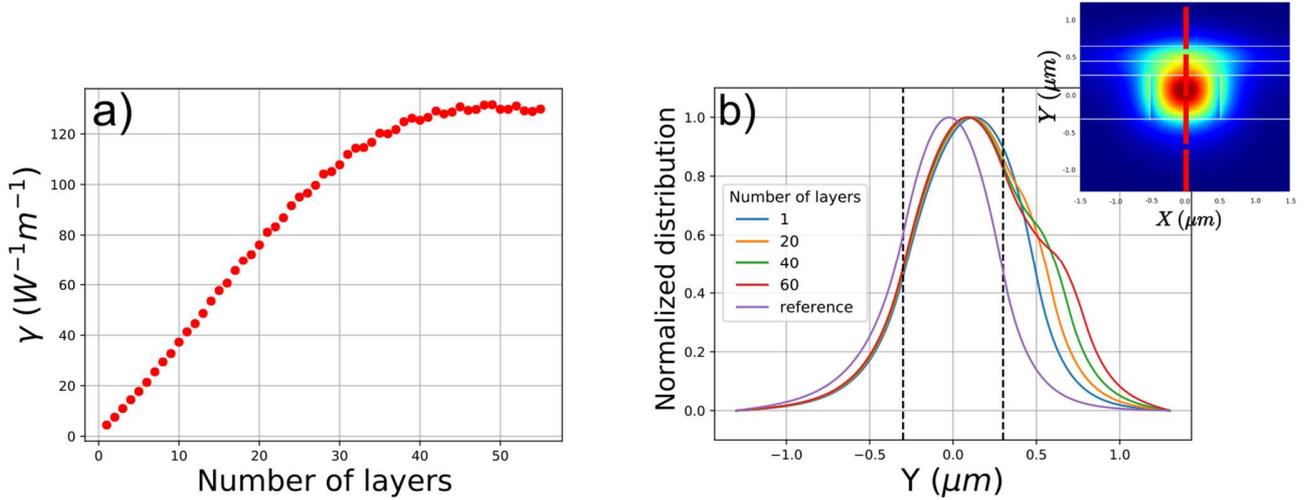

**Fig. 4. a) Effective nonlinear coefficient (γ) versus the number of layers. b) Slice of the transverse profile of the TE mode taken along the dashed red line (see inset). The dashed black lines represent the core boundaries. The reference corresponds to a standard planarized buried SiN waveguide without anything on top.**

In Fig 3a), we have the nonlinear coefficient $\gamma$ calculated with eq. 6 for varying cross section dimensions: a 5 times increase of $\gamma$ is observed, $\gamma$ being a function of the cross-section dimensions. The same trend can be observed in Fig 3b) with the overlap between the $MoS_2$ and the TE mode. Comparing Fig 3b) and Fig 3c) suggests that the $\gamma$ increase is directly related to the mode/TMD layer overlap. Obviously, the number of layers can thus be further increased, which is also based on very convincing feasibility elements with regard to the possibility of stacking a series of 2D material sheets [20]. The related results are displayed in Fig 4. Again, on top of the SiN/$MoS_2$, a 200 nm layer of $HfO_2$ is considered in order to attract the mode upward. This introduces an impressive boosting of the effective nonlinear coefficient $\gamma$ with an increase of about 2 orders of magnitudes for the maximum value when compared with a standard SiN planarized buried strip. A saturation of the improvement above a certain number of layers is also observed (e.g. 40). At some point, the optical mode spreads too much, thus causing a decrease of the field density around the 2D material layers. However, this limitation sets a very comfortable limit to the optimization of the hybrid TMD-SiN waveguides, which nonlinear coefficients can reach values of more than 100 $W^{-1}$ $m^{-1}$, thus almost approaching nonlinear coefficients of silicon waveguides [12].

In the continuity, we draw our interest in an emblematic application with the enhanced waveguide nonlinearity. Our choice fell on the wavelength conversion mechanism by degenerate FWM in a ring resonator structure based on the previously optimized hybrid waveguides. The device is an all pass resonator device as presented Fig 5b) [33]. For this study, we relied on a formalism developed and described in several previous papers [20,34,35]. We recap here the main ideas of the considered formalism starting with the conversion efficiency in a straight waveguide [34,35]:

$$\eta_{straight} = \frac{P_C(L)}{P_S(0)} = |\gamma P_0 L_{eff}|^2 exp(-\alpha L) \quad (6)$$

, with $\alpha$ are the propagation losses, $L$ the waveguide propagation length, $P_0$ is the pump power, $P_C$ is the power of the converted (idler) frequency, $P_S$ the power of the signal frequency (as presented Fig. 5b)) and $L_{eff}$ is the effective propagation length defined as:

$$L_{eff} = \frac{1 - e^{-(\alpha + j\kappa)L}}{\alpha + j\kappa} \quad (7)$$

, $\kappa$ being the phase mismatch. Now, considering a ring resonator and 3 frequencies, $\omega_p$, $\omega_s$ and $\omega_c$ i.e. the pump, the signal and the converted (idler) respectively, the feedback due to the cavity gives rise to the appearance of Field Enhancement coefficients ($FE_i$) in the conversion efficiency:

$$\eta_{resonator} = \frac{P_C(L)}{P_S(0)} = |\gamma P_0 L_{eff}|^2 FE_p^4 FE_c^2 FE_s^2 exp(-\alpha L) \quad (8)$$

, with $k$ and $r$ the cross and pass coefficients of the coupling area respectively:

$$FE_i = \left| \frac{k}{1 - re^{\frac{-\alpha L}{2} + j\frac{\omega_i L}{c}}} \right| \quad (9)$$

This conversion efficiency was made for two devices with identical cross sections but different settings: one for a straight waveguide and the other for a regular micro-ring with a 100 μm radius, the considered unfolded propagation lengths being set to the same values to enable the comparison between the designs. Those results are presented in Fig 5 for degenerate FWM processes around 1.55μm wavelength. The considered pump power in this study is stated as 5 mW, a very reasonably low power level for a continuous wave pumping optical source. The calculations were performed for frequencies at neighboring resonances (see Fig 5a)). The comparison was made using different parameters as variables. The different types of waveguide designs translate into different effective nonlinear $\gamma$ factors. The $\gamma \approx 1$ W$^{-1}$m$^{-1}$ case corresponds to a standard reference SiN buried strip planarized waveguide without further active material. The three other values are for the low, middle ground, and high values of the showcased structure of Fig.3. What appears clearly is that a strong transmission coefficient is required from the coupling area of the waveguide in order to make the effect of the ring interesting. This is expected when considering that a high-quality factor can be linked to high photon lifetime in the cavity. Moreover, for a ring with a high coupling transmission, the conversion efficiency remains higher for the ring than for the straight waveguide even for very high losses. However, as the propagation losses increase, the overall quality factor is reduced, hence reducing the conversion efficiency faster than in a straight waveguide. In any case, using a proper design (high coupling transmission) allows a conversion efficiency as high as -6.3 dB conversion efficiency for an injected input power of 1mW ($P_L(0)$=1mW) (see Fig 5), which compares well to similar studies conducted in the SOI platform [35].

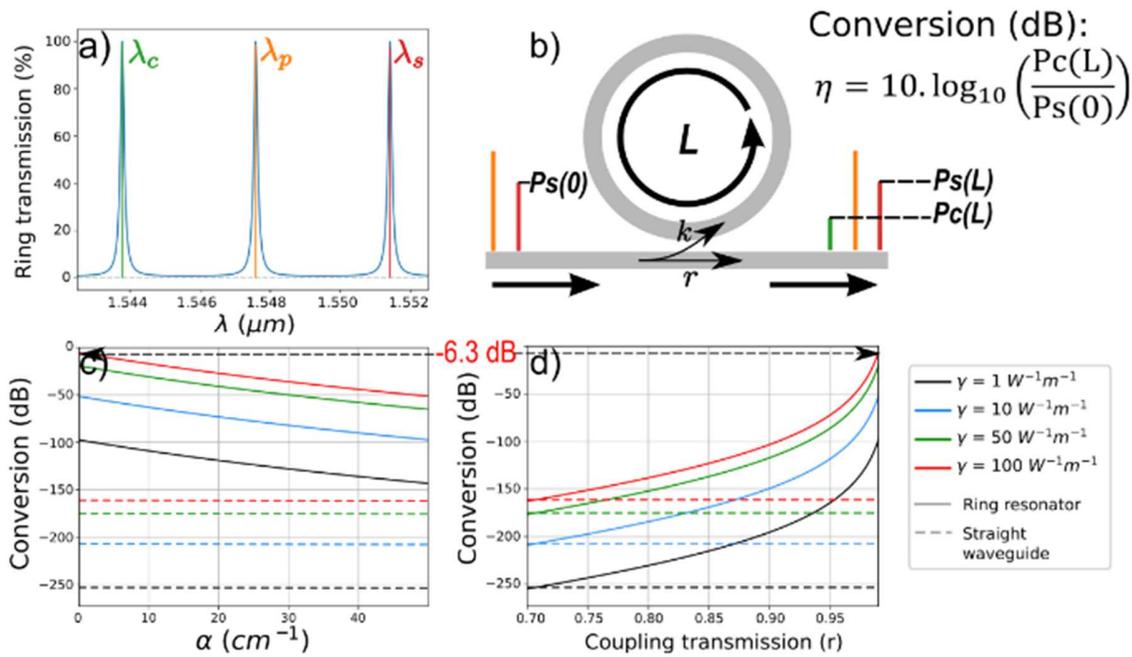

**Fig. 5.a) Transfer function of a 100 μm radius ring resonator around 1.55 μm. b) Ring resonator coupling area. $k$ and $r$ are the cross and pass coefficients respectively. $P_S(0), P_S(L), P_C(L)$ are respectively the input signal power, the output signal power and the output converted (idler) power. c),d) Conversion efficiency $\eta$ in dB for straight waveguides (dashed lines) and rings (full lines). In c) $\eta$ is plotted versus the propagation losses (for $r$=0.99) and in d), $\eta$ is plotted versus the coupling transmission coefficient $r$ (for $\alpha$=0.7 cm$^{-1}$)**

To conclude, we propose a novel approach for boosting of the nonlinear photonic SiN/SiO$_2$ platform thanks to the proper integration of MoS$_2$ monolayers in SiN planarized waveguides. The strategy relies on dispersion engineering and optimizing of the waveguide structure to maximize the interaction between the TE mode and MoS$_2$. Using multiple mono-layers of MoS$_2$ separated by oxide spacers while pulling the mode vertically with a top cladding high index dielectric layer, we design hybrid photonic waveguides with nonlinear effective coefficients up to $\gamma$~120 W$^{-1}$m$^{-1}$. Micro-ring resonators relying on these optimized nonlinear waveguides show potentials for wavelength conversion efficiencies reaching -6.3 dB for a pump power levels as low as 5 mW at 1.55µm wavelength. We believe that the proposed approach will provide useful guidelines for further development in integrated nonlinear optics and its applications such as frequency combs, supercontinuum and quantum light sources relying on the integration of 2D materials in the silicon nitride waveguide platform.

**Disclosures.** The authors declare no conflicts of interest.

**Acknowledgments.** This work was supported by Université Paris Saclay within the Centre for nanoscience and nanotechnology (C2N) in France and Aalto University in Finland.